\documentclass[aps,reprint,superscriptaddress]{revtex4-1}

\pdfoutput=1

\usepackage{graphics,epsfig}
\usepackage{amsmath,amssymb}
\usepackage{bm}
\usepackage{subfigure}

\begin{document}

\title{Phase ordering, transformation, and grain growth of two-dimensional binary
colloidal crystals: A phase field crystal modeling}

\author{Doaa Taha}
\affiliation{Department of Physics and Astronomy, Wayne State University,
Detroit, Michigan 48201, USA}
\author{S. R. Dlamini}
\affiliation{Department of Physics and Astronomy, Wayne State University,
Detroit, Michigan 48201, USA}
\author{S. K. Mkhonta}
\affiliation{Department of Physics, University of Eswatini, %Private Bag 4,
Kwaluseni M201, Eswatini}
\author{K. R. Elder}
\affiliation{Department of Physics, Oakland University, Rochester, Michigan 48309, USA}
\author{Zhi-Feng Huang}
\affiliation{Department of Physics and Astronomy, Wayne State University,
Detroit, Michigan 48201, USA}

\date{\today}

\begin{abstract}

The formation and dynamics of a wide variety of binary two-dimensional ordered structures
and superlattices are investigated through a phase field crystal model with sublattice
ordering. Various types of binary ordered phases, the phase diagrams, and the grain growth
dynamics and structural transformation processes, including the emergence of topological
defects, are examined.  The results are compared to the ordering and assembly of
two-component colloidal systems. Two factors governing the binary phase ordering are
identified, the coupling and competition between the length scales of two sublattices
and the selection of average particle densities of two components. The control and
variation of these two factors lead to the prediction of various complex binary ordered
patterns, with different types of sublattice ordering for integer vs. noninteger ratios
of sublattice length scales. These findings will enable further systematic studies of
complex ordering and assembly processes of binary systems particularly binary colloidal
crystals.
  
\end{abstract}

\maketitle

\section{Introduction}

The assembly of binary colloidal crystals (BiCCs) with sublattice ordering has been
of significant interest in various aspects of fundamental research and applications
\cite{Nat_Gold1998,KhalilNatCommun12,KimAdvMater05,Redl2003_metamaterial,
Shevchenko2006_metamaterial,Broadband_Cai_acsami,LiqCrys_Ognysta_Lang2009,Honold_BH_2016}. 
These artificial ordered systems can be synthesized from diverse types of building
components that vary in size and shape and are selected or tailored to possess
specific functionalities \cite{Redl2003_metamaterial,Shevchenko2006_metamaterial}.
Variations of spatial arrangement, shape, and size of structural components translate
into different macroscopic properties of the system and the fabrication of the
corresponding functional materials (e.g., photonics \cite{PhotonicCrystal_Wan2009}
and semiconductors \cite{semiconduc_Evers_nl,semiconduc_chen_acsn}), in addition to
biological applications (e.g., cell culture substrates \cite{CellCulture_WangJOMCB,
CellCulture_Wang2016} and MRI contrast agents \cite{MRI_Kostiainen2013}). Although
a great deal of effort has been placed on the synthesis of colloidal systems 
from self or directed assembly of colloidal particles or building blocks, it still
remains a challenging task to precisely control and predict the structural and
dynamic properties of the system. Many system parameters and growth or processing
conditions, such as entropy \cite{Entropy_Eldridge_Nat}, temperature \cite{tempdep_JACS},
external magnetic or electric fields \cite{KhalilNatCommun12,ACBinaryHeatley},
isotropic and anisotropic interparticle interactions, and system elasticity and
plasticity, determine the structural diversity of the assembly. 

One of the key challenges for understanding the complex phenomena associated with
colloid assembly is the development of theoretical approaches that can efficiently model
nonequilibrium phenomena with multiple length scales and diffusive time scales for
large enough systems of experimental relevance. Various theoretical methods have been
developed to study binary colloidal structures and the associated phenomena. For example,
Monte Carlo (MC) simulations have been conducted to examine the structure factor of
charged binary colloidal mixtures \cite{Petris2003} as well as the phase transformation
in a two-dimensional (2D) BiCC monolayer that is consistent with experiments
\cite{YangSoftMatter15}. Similarly, molecular dynamics (MD) simulations have predicted
that 2D BiCCs can only be achieved for certain particle ratios \cite{Stirner2005}.
However, the system size and time range are usually limited in these atomistic
simulations, given the large computational demands to access large-scale behaviors
of the system. 

Recently progress has been made to overcome these limitations by the development
of multiple scale approaches. Among them is the phase field crystal (PFC) method
\cite{Elder02,*Elder04} that introduces crystalline ordering into the traditional
phase-field type continuum approach. PFC models, motivated from the classical
density functional theory (DFT) of freezing \cite{elder07,huang10b}, incorporate
the small length scales of crystalline materials (including the basic features of
the crystalline state such as elasticity, plasticity, defects, and multiple crystal
orientations) on diffusive time scales. The system evolution is governed by
dissipative and relaxational dynamics driven by free energy minimization. This
method thus bridges the gap between continuum modeling that describes the long
wavelength behavior of the system but not crystalline details, and atomistic
modeling that captures the microscopic details but is computationally challenging for
large systems. It has been successfully applied to the study of a broad range of
phenomena such as quantum dot growth during epitaxy \cite{HuangEpi2008,*HuangEpi2010},
grain boundaries of 2D materials \cite{Petri_Graphen2016,Taha17}, graphene Moir\'e
patterns \cite{Marie_GrapMoire_2017}, colloidal solidification and growth
\cite{Teeffelen_ColSold2009,TegzeSoftMatter11}, structural phase transformation
\cite{greenwood2011modeling,Ofori-opoku13}, glass formation \cite{Berry_Glass2008},
and quasicrystal growth \cite{Achim14}, among many others. For the application of
PFC method to colloidal systems, most existing studies are limited to single-component
crystallization process \cite{Teeffelen_ColSold2009,TegzeSoftMatter11}, while the
study of binary colloidal structures is still lacking.

In this paper, we extend the PFC method to study various types of binary 2D colloidal
structures with sublattice ordering, based on a binary PFC model developed in our
prior work \cite{Taha17}. We start by deriving the model from classical DFT for
a two-component system, keeping only two- and three-point interparticle direct
correlations. The ordered structures of BiCC are found to be determined by the
coupling between different sublattice length scales, as well as the average density
variations of the particle species. For equal sublattice length scales we identify
seven binary phases, with results consistent with recent experimental findings
\cite{Nat_Gold1998,KhalilNatCommun12,KimAdvMater05,Honold_BH_2016,YangSoftMatter15,
SteinNatCommun16}. Using analytic and numerical methods the stability of various
phases and the coexistence between them are determined and used to construct phase
diagrams. In addition, numerical simulations are employed to examine the dynamical
processes of grain growth and phase transformation for different binary ordered
structures, including the formation of various types of topological defects during
the system evolution. Importantly, varying the length scale ratio between the two
sublattices allows us to access and predict a much broader range of complex ordered
(or quasicrystalline) patterns and superlattices, with results depending on integer
vs. noninteger type of the ratio and the choice of densities of the two components.

\section{Model derivation}

The PFC equations for binary AB system can be derived from classical dynamic 
density functional theory (DDFT), following the procedure described in
Ref. \cite{huang10b}. In classical DFT the free energy functional
for a two-component system is expanded as (see, e.g., Ref. \cite{re:singh91})
\begin{eqnarray}
&& F/k_BT = \int d{\bm r} \sum_{i} \left [ \rho_i \ln
    \left ({\rho_i }/{\rho_l^i} \right ) - \delta \rho_i \right ]
  - \sum_{n=2}^{\infty}\frac{1}{n!} \label{eq:F_DFT_AB}\\
&&  \times \int d{\bm r_1} \cdots d{\bm r_n}
\sum_{i,...,j} C_{i...j}^{(n)}({\bm r_1}, \cdots, {\bm r_n}) \delta
\rho_i({\bm r_1}) \cdots \delta \rho_j ({\bm r_n}), \nonumber
\end{eqnarray}
where $\delta \rho_i = \rho_i - \rho_l^i$, $\rho_{i=A,B}$ is the local atomic
number density of A, B component, $\rho_l^i$ is a reference-state density of
$i$ component, and $C_{i...j}^{(n)}$ is the $n$-point direct correlation
function between $i,...,j=A, B$. The dynamics of density fields is governed
by the DDFT equations \cite{re:marconi99,re:archer05a,huang10b}
\begin{eqnarray}
&&{\partial \rho_A}/{\partial t} = {\bm \nabla} \cdot \left ( 
  M_A \rho_A {\bm \nabla} \frac{\delta F}{\delta \rho_A}
  + \sqrt{\rho_A} {\bm \eta}_A \right ), \nonumber\\
&&{\partial \rho_B}/{\partial t} = {\bm \nabla} \cdot \left ( 
  M_B \rho_B {\bm \nabla} \frac{\delta F}{\delta \rho_B}
  + \sqrt{\rho_B} {\bm \eta}_B \right ),
\label{eq:ddft_AB}
\end{eqnarray}
where $M_{A(B)}$ is the mobility of A(B) component and ${\bm \eta}_{A(B)}$
is the noise field. In principle the free energy in classical DFT contains
all the effects including noise, since the DFT derivation comes from the
partition function summing over all states at some finite temperature and
the resulting equilibrium $F$ is a functional of noise-averaged densities
(i.e., after ensemble average). Adding an extra noise term in DDFT would
then result in a double counting of fluctuations \cite{re:marconi99}.
From a more pragmatic point of view, DDFT without fluctuations however,
misses key dynamic process, such as nucleation events. In practice the
free energy functional used and the corresponding density fields are usually
coarse-grained, for which the governing DDFT equations should be stochastic
as demonstrated in Ref.~\cite{re:archer04b}. Therefore, for completeness
noise is incorporated in the above dynamic equations, where $F$ should
then be considered as an effective, coarse-grained free energy functional
but not the true free energy.

Defining the density variation fields $n_A=(\rho_A-\rho_l^A)/\rho_l$ and
$n_B=(\rho_B-\rho_l^B)/\rho_l$ (with $\rho_l=\rho_l^A+\rho_l^B$), keeping only
two- and three-point direct correlations $C_{ij}^{(2)}({\bm r_1}, {\bm r_2})$
and $C_{ijk}^{(3)}({\bm r_1},{\bm r_2}, {\bm r_3})$ ($i,j,k=A,B$), and expanding
them in Fourier space (with wave number $q$) via
\begin{eqnarray}
  &&\hat{C}_{ij}^{(2)}(q)=-\hat{C}_0^{ij}+\hat{C}_2^{ij} q^2-\hat{C}_4^{ij} q^4
  +\cdots, \nonumber\\
  &&\hat{C}_{ijk}^{(3)}({\bm q},{\bm q'}) \simeq
  \hat{C}_{ijk}^{(3)}({\bm q}={\bm q'}=0)=-\hat{C}_0^{ijk}, \label{eq:Cq}
\end{eqnarray}
we can rewrite Eq. (\ref{eq:F_DFT_AB}) as
\begin{eqnarray}
\frac{\Delta F}{\rho_l k_BT} =&& \int d{\bm r} \left \{
\Delta \rho_l^A \left ( 1+\frac{n_A}{\Delta \rho_l^A} \right ) 
\ln \left ( 1+\frac{n_A}{\Delta \rho_l^A} \right ) - n_A \right. \nonumber\\
&& + \Delta \rho_l^B \left ( 1+\frac{n_B}{\Delta \rho_l^B} \right ) 
\ln \left ( 1+\frac{n_B}{\Delta \rho_l^B} \right ) - n_B \nonumber\\
&& + \frac{\rho_l}{2} \left [ \hat{C}_0^{AA} n_A^2 + n_A \left ( 
  \hat{C}_2^{AA} \nabla^2 + \hat{C}_4^{AA} \nabla^4 \right ) n_A \right. \nonumber\\
&& + \hat{C}_0^{BB} n_B^2 + n_B \left (\hat{C}_2^{BB} \nabla^2 +
  \hat{C}_4^{BB} \nabla^4 \right ) n_B \nonumber\\
&& \left. + 2\hat{C}_0^{AB} n_An_B + 2n_A \left ( \hat{C}_2^{AB} \nabla^2 +
  \hat{C}_4^{AB} \nabla^4 \right ) n_B \right ] \nonumber\\
&& + \frac{\rho_l^2}{6} \left [ \hat{C}_0^{AAA} n_A^3
  + \hat{C}_0^{BBB} n_B^3 \right. \nonumber\\
&& \left. \left. + 3\hat{C}_0^{AAB} n_A^2n_B + 3\hat{C}_0^{ABB} n_An_B^2
  \right ] \right \},
\label{eq:F_nAB}
\end{eqnarray}
which is the same as Eq. (A1) in the appendix of Ref.~\cite{huang10b},
with $\Delta \rho_l^{A(B)} = \rho_l^{A(B)}/\rho_l$. Note that
Eqs. (\ref{eq:Cq}) and (\ref{eq:F_nAB}) are based on the assumption of
only one characteristic length scale for either A or B sublattice (as
determined by $\hat{C}_{ij}^{(2)}$), and the approximation of $\hat{C}_{ijk}^{(3)}$
only by its zero-wavevector component as used in previous classical DFT work
for hard-spheres \cite{re:smithline87} and Lennard-Jones \cite{re:rick89}
binary systems.

Substituting Eq. (\ref{eq:F_nAB}) into the DDFT Eqs. (\ref{eq:ddft_AB}), choosing
the same reference state for A and B, i.e., $\rho_l^A = \rho_l^B$, and keeping
only the leading order terms (via scale analysis), we can derive a new binary
PFC model represented by
\begin{eqnarray}
&& \partial n_A / \partial t = D_A \nabla^2 \frac{\delta \mathcal{F}}{\delta n_A}
  + {\bm \nabla} \cdot {\bm \eta}_A, \nonumber\\
&& \partial n_B / \partial t = D_B \nabla^2 \frac{\delta \mathcal{F}}{\delta n_B}
  + {\bm \nabla} \cdot {\bm \eta}_B,
\label{eq:nA_nB_}
\end{eqnarray}
where the diffusion coefficients $D_{A(B)}=M_{A(B)}k_BT$, and the resulting PFC
free energy functional is given by 
\begin{eqnarray}
  &\mathcal{F} = & \int d{\bm r}  \left [ \frac{1}{2} \Delta B_A n_A^2
    + \frac{1}{2} B_A^x n_A \left ( R_A^2 \nabla^2 + 1 \right )^2 n_A \right. \nonumber\\
  && +\frac{1}{2} \Delta B_B n_B^2  + \frac{1}{2} B_B^x n_B
    \left ( R_B^2 \nabla^2 + 1 \right )^2 n_B \nonumber\\
  && - \frac{1}{3} \tau_A n_A^3 + \frac{1}{4} v_A n_A^4
    - \frac{1}{3} \tau_B n_B^3 + \frac{1}{4} v_B n_B^4 \nonumber\\
  && + \Delta B_{AB} n_A n_B + B_{AB}^x n_A \left ( R_{AB}^2 \nabla^2 + 1 \right )^2 n_B
    \nonumber\\
  && \left. +\frac{1}{2} w_0 n_A^2 n_B + \frac{1}{2} u_0 n_A n_B^2 \right ].
  \label{eq:F_}
\end{eqnarray}
Here $\Delta B_{A(B)} = B_{A(B)}^l - B_{A(B)}^x$, $\Delta B_{AB} = B_{AB}^l - B_{AB}^x$,
and all the parameters can be expressed via expansion coefficients of two- and
three-point direct correlation functions in Fourier space, i.e.,
\begin{eqnarray}
  && B_A^x = \frac{\rho_l^A \left.\hat{C}_2^{AA}\right.^2}{4\hat{C}_4^{AA}}, \;\;
  B_A^l = 1+\rho_l^A\hat{C}_0^{AA}, \;\;
  R_A = \sqrt{\frac{2\hat{C}_4^{AA}}{\hat{C}_2^{AA}}}, \nonumber\\
  && \tau_A = -\frac{\rho_l}{2} \left ( \hat{C}_0^{AA} + \rho_l^A \hat{C}_0^{AAA} \right ),
  \;\; v_A = \frac{\rho_l^2}{3}\hat{C}_0^{AAA}, \nonumber\\
  && B_B^x = \frac{\rho_l^B \left.\hat{C}_2^{BB}\right.^2}{4\hat{C}_4^{BB}}, \;\;
  B_B^l = 1+\rho_l^B\hat{C}_0^{BB}, \;\;
  R_B = \sqrt{\frac{2\hat{C}_4^{BB}}{\hat{C}_2^{BB}}}, \nonumber\\
  && \tau_B = -\frac{\rho_l}{2} \left ( \hat{C}_0^{BB} + \rho_l^B \hat{C}_0^{BBB} \right ), 
  \quad v_B = \frac{\rho_l^2}{3}\hat{C}_0^{BBB}, \nonumber\\
  && B_{AB}^x = \frac{\rho_l^A \left.\hat{C}_2^{AB}\right.^2}{4\hat{C}_4^{AB}}, \;\;
  B_{AB}^l = \rho_l^A\hat{C}_0^{AB}, \;\;
  R_{AB} = \sqrt{\frac{2\hat{C}_4^{AB}}{\hat{C}_2^{AB}}}, \nonumber\\
  && w_0 = \rho_l \rho_l^A \hat{C}_0^{AAB}, \;\; u_0 = \rho_l \rho_l^A \hat{C}_0^{ABB}.
\end{eqnarray}
To reduce the number of parameters we can rescale the above PFC equations in terms
of A parameters, i.e., via a length scale $R_A$, a time scale $R_A^2/(D_A B_A^x)$,
and $n_{A(B)} \rightarrow n_{A(B)} \sqrt{v_A/B_A^x}$, leading to
\begin{equation}
  \frac{\partial n_A}{\partial t} = \nabla^2 \frac{\delta \mathcal{F}}{\delta n_A}
  + {\bm \nabla} \cdot {\bm \eta}_A, \quad
  \frac{\partial n_B}{\partial t} = m_B \nabla^2 \frac{\delta \mathcal{F}}{\delta n_B}
  + {\bm \nabla} \cdot {\bm \eta}_B,
\label{eq:nA_nB}
\end{equation}
where $m_B=M_B/M_A$ represents a mobility contrast between A and B species, and
the rescaled noise fields satisfy the conditions
\begin{eqnarray}
&\langle {\bm \eta}_A \rangle = \langle {\bm \eta}_B \rangle 
= \langle {\bm \eta}_A {\bm \eta}_B \rangle = 0,& \nonumber\\
&\langle \eta_i^{\mu}({\bm r},t) \eta_i^{\nu}({\bm r}',t) \rangle 
= 2 \Gamma_i k_B T \delta({\bm r}-{\bm r}') \delta(t-t') \delta^{\mu\nu},&
\end{eqnarray}
with $i=A, B$, $\mu, \nu = x, y$ for a 2D system, and the rescaled noise amplitudes
$\Gamma_B / \Gamma_A = M_B / M_A = m_B$. The PFC free energy functional is rescaled as
\begin{eqnarray}
  &\mathcal{F} = & \int d{\bm r}  \left [ -\frac{1}{2} \epsilon_A n_A^2
    + \frac{1}{2} n_A \left ( \nabla^2 + q_A^2 \right )^2 n_A \right. \nonumber\\
  && -\frac{1}{2} \epsilon_B n_B^2  + \frac{1}{2} \beta_B n_B
    \left ( \nabla^2 + q_B^2 \right )^2 n_B \nonumber\\
  && - \frac{1}{3} g_A n_A^3 + \frac{1}{4} n_A^4
    - \frac{1}{3} g_B n_B^3 + \frac{1}{4} v n_B^4 \nonumber\\
  && + \alpha_{AB} n_A n_B + \beta_{AB} n_A
    \left ( \nabla^2 + q_{AB}^2 \right )^2 n_B \nonumber\\
  && \left.  +\frac{1}{2} w n_A^2 n_B + \frac{1}{2} u n_A n_B^2 \right ],
  \label{eq:F}
\end{eqnarray}
where the dimensionless parameters are given by:
$q_A=1$ (due to rescaling), $q_B=R_A/R_B$, $q_{AB}=R_A/R_{AB}$,
$\epsilon_{A(B)} = -\Delta B_{A(B)} / B_A^x = (B_{A(B)}^x - B_{A(B)}^l) / B_A^x$,
$\alpha_{AB} = \Delta B_{AB} / B_A^x$, $\beta_{AB} = B_{AB}^x/(B_A^x q_{AB}^4)$,
$\beta_B = B_B^x/(B_A^x q_B^4)$, $g_{A(B)} = \tau_{A(B)} / \sqrt{B_A^x v_A}$,
$v=v_B/v_A$, $w = w_0 / \sqrt{B_A^x v_A}$, and $u = u_0 / \sqrt{B_A^x v_A}$.

\section{Ordered structures and phase diagrams: Equal length scales}

\begin{figure}
\centerline{\includegraphics[width=0.5\textwidth]{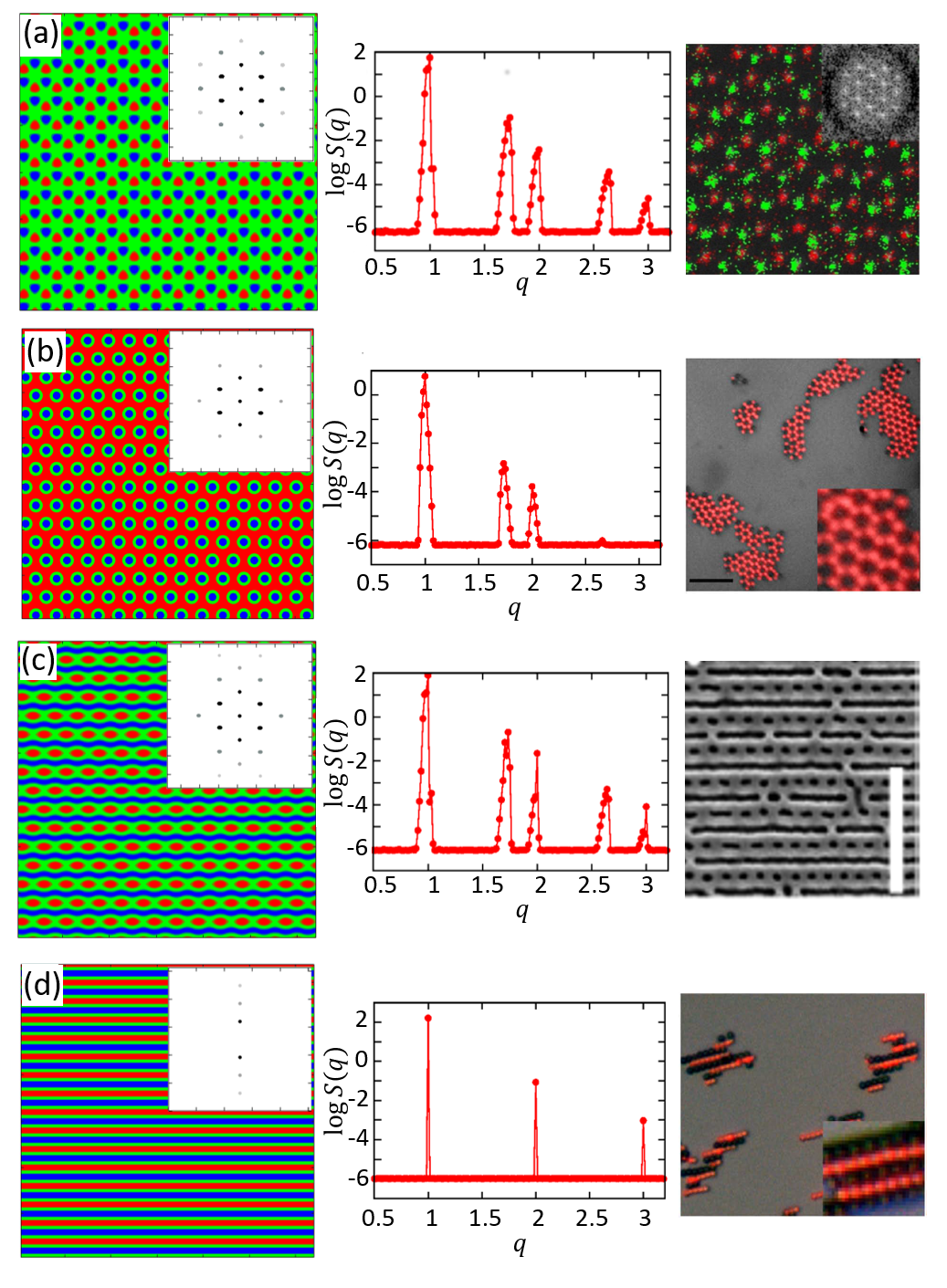}}
\caption{Some ordered phases obtained from PFC simulations for the case of equal
  sublattice length scales ($q_A=q_B=1$), including (a) binary honeycomb (BH),
  (b) triangular B \& honeycomb A (TBHA), (c) elongated triangular A \& stripe B
  (ETASB), and (d) binary stripe (BS). For the first-column simulation results
  the red-colored locations correspond to the maximum density of component A,
  while the blue-colored ones to the maximum density of B. The corresponding
  diffraction patterns are shown as insets and the circularly averaged structure
  factors are given in the second column. As a comparison the third column shows
  the related results observed in previous experiments, reprinted with permission
  from Ref.~\cite{Honold_BH_2016} in (a), from Ref.~\cite{KhalilNatCommun12} in
  (b) and (d), and from Ref.~\cite{SteinNatCommun16} in (c).}
  \label{fig:phases}
\end{figure}

\begin{figure}
\centerline{\includegraphics[width=0.4\textwidth]{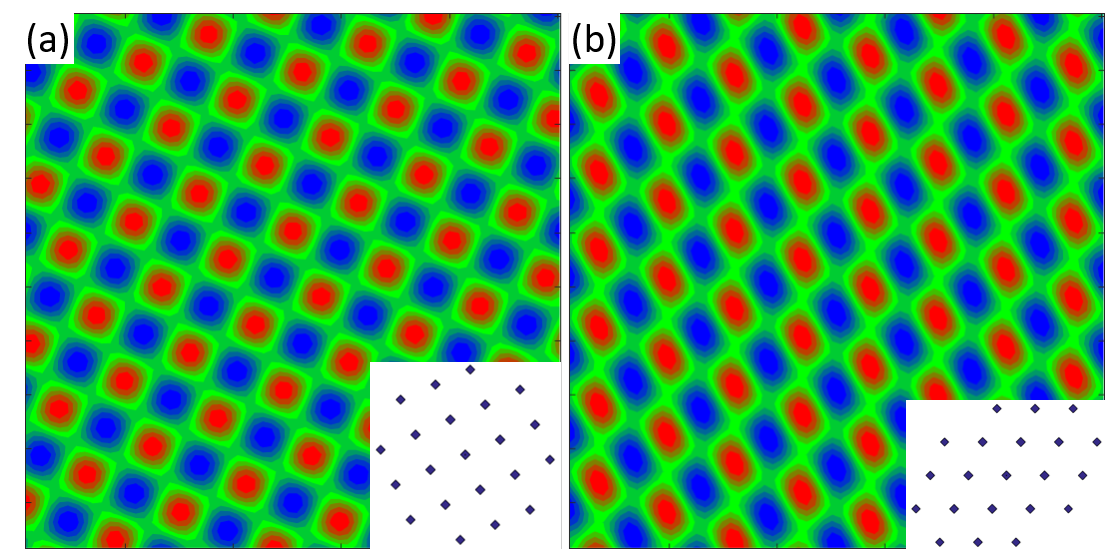}}
\caption{(a) Binary square (BSq; i.e., checkerboard) and (b) binary rhombic (BR)
  phases obtained from PFC simulations, with $q_A=q_B=1$ and positive average
  densities of A and B components [$n_{A0}=0.53$, $n_{B0}=0.82$ for (a) and
  $n_{A0}=0.55$, $n_{B0}=0.80$ for (b)]. The color scheme is the same as that of
  Fig.~\ref{fig:phases}. Each inset shows the corresponding diffraction pattern.}
  \label{fig:phases2}
\end{figure}

The binary PFC model constructed here [i.e., Eqs. (\ref{eq:nA_nB})--(\ref{eq:F})],
although only including one mode for each of the sublattices, can produce a rich
variety of ordered structures as well as their coexistence. Detailed results
depend on the selection and competition of length scales between the two
sublattices. For simplicity, in this section we consider the case of equal
lattice spacing of A and B sublattices and zero mobility contrast, such that
$q_A=q_B=q_{AB}=1$ and $m_B=1$, and use the model parameters of $\alpha_{AB}=0.5$,
$\beta_{AB}=0.02$, $g_A=g_B=0.5$, $w=u=0.3$, and $\beta_B=v=1$. For these
parameters a total of seven stable phases of 2D binary sublattice ordering have
been identified, with some structures and the corresponding diffraction patterns
and/or circularly averaged structure factors for density difference $n_A-n_B$
shown in Figs. \ref{fig:phases} and \ref{fig:phases2}. These binary structures
or superlattices are basically the combinations of triangular, stripe, inverse
triangular (noting that an inverse triangular lattice is of honeycomb structure),
square, rhombic, and homogeneous states of A and B sublattices, and are
determined by the coupling between $n_A$ and $n_B$ density fields. They include
(i) a binary honeycomb (BH) phase with triangular A and B sublattices, (ii) a binary
stripe (BS) phase with A and B stripe sublattices, (iii) a combination of elongated
triangular A (or B) sublattice and stripe B (or A) sublattice (ETASB or ETBSA),
(iv) a pattern with triangular A(B) sublattice but inverse triangular (i.e., honeycomb)
structure of B(A) sublattice (TAHB or TBHA), and (v) a binary homogeneous (BHom) state.
In addition, two other ordered phases can be found only from the regime of positive
average density variations, including (vi) a checkerboard structure showing as binary
square (BSq) sublattices of A and B, and (vii) a binary rhombic (BR) phase consisting
of A and B rhombic sublattices (see Fig.~\ref{fig:phases2}). These seven phases are
identified through our numerical simulations of the dynamic Eq.~(\ref{eq:nA_nB}),
across different ranges of average density values including $n_{A0}$, $n_{B0}$ varying
from $-0.5$ to $0.5$ to obtain phases (i)--(v) and from $0.5$ to $1$ to obtain phases
(vi) and (vii). It is possible that more ordered phases could be found across a
broader range of parameter space as a result of the nonlinear coupling between A
and B sublattice density fields.

The TAHB (or TBHA) phase has been observed in experiments of 2D binary
colloid mixtures \cite{Nat_Gold1998,KhalilNatCommun12,KimAdvMater05}, while
the BSq structure has been achieved in both experiments and MC simulations
of binary colloidal monolayers \cite{KhalilNatCommun12,YangSoftMatter15}.
The BH phase not only has been obtained in previous experiments of binary
colloidal system with honeycomb symmetry \cite{Honold_BH_2016}, but also
corresponds to the structure of binary 2D hexagonal materials such as hexagonal
boron nitride (\textit{h}-BN). The BS structure has also been observed in binary
colloids, although the stripe phase obtained in our modeling would be more
relevant to that of diblock copolymers given the homogeneous density distribution
within each stripe. Although to the best of our knowledge the ETASB (or ETBSA)
superlattice has not been found in colloidal systems, a similar phase has been
produced in thin film experiments of binary blends of block copolymers
(controlled by substrate surface pre-patterning) \cite{SteinNatCommun16}.
Some of the corresponding experimental images are shown in Fig.~\ref{fig:phases}
for comparison.

\subsection{Phase diagrams: Analytics from one-mode approximation}
\label{sec:analytic}

\begin{figure*}
  \centerline{\includegraphics[width=0.8\textwidth]{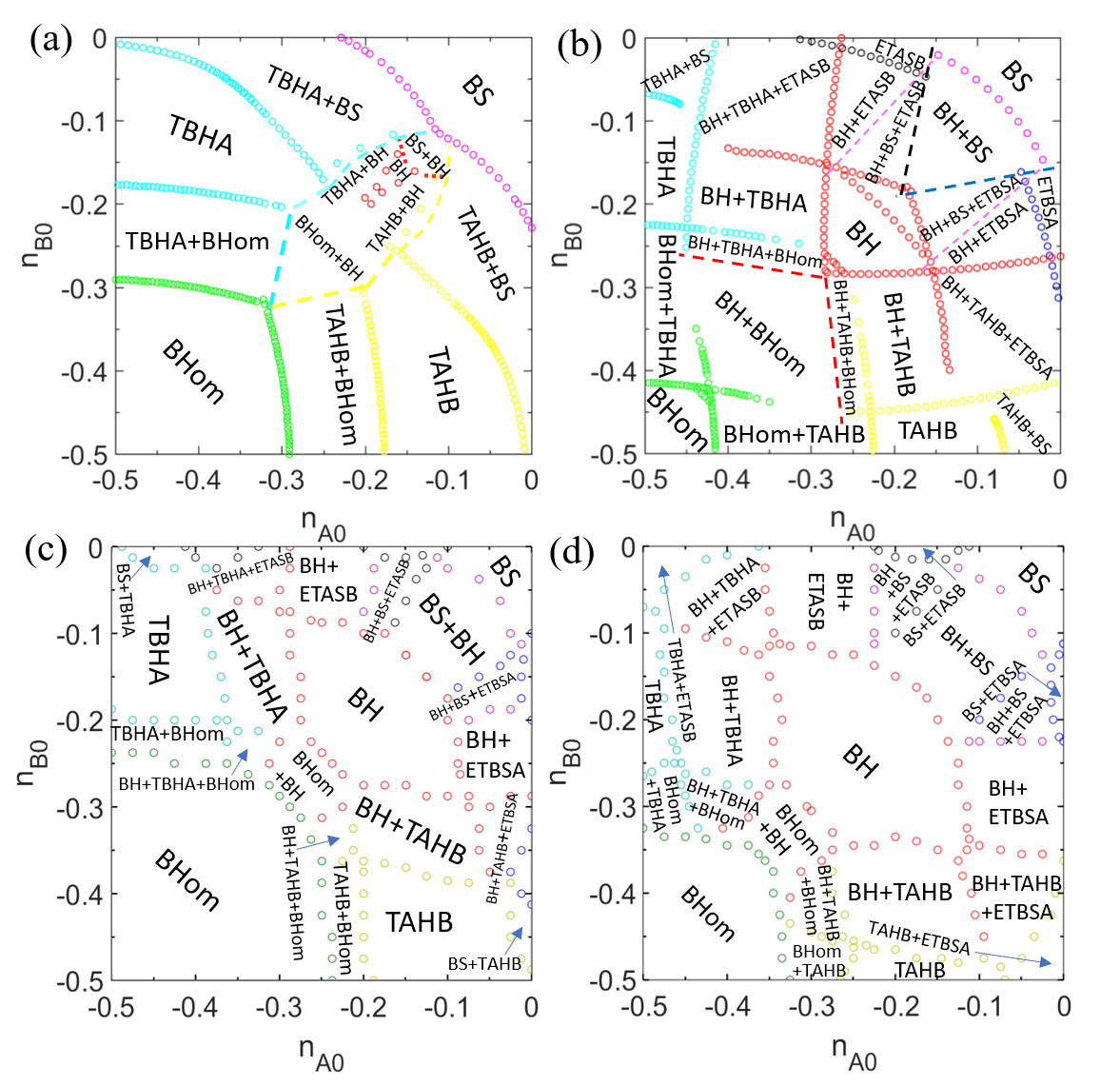}}
  \caption{Phase diagrams of the binary PFC model in the cross-section plane of
    $n_{A0}$ vs. $n_{B0}$ (for $n_{A0}, n_{B0} <0$), for the case of equal sublattice
    length scales $q_A=q_B=1$ at $\epsilon_A=\epsilon_B=0.1$ [(a) and (c)] and $0.3$
    [(b) and (d)]. Results of the analytic calculations are shown in (a) and (b),
    while those determined by direct numerical simulations are given in (c) and (d).}
  \label{fig:phase_diagrams}
\end{figure*}

The corresponding phase diagram of this PFC model can be determined via standard
thermodynamics. The phase boundaries for the coexistence between any two phases
1 and 2 are calculated by the conditions
\begin{equation}
\mu_{A1} = \mu_{A2}, \qquad \mu_{B1} = \mu_{B2}, \qquad \omega_1 = \omega_2,
\label{eq:coexist}
\end{equation}
where $\mu_{A(B)} = \partial f / \partial n_{A(B)0}$ is the chemical potential for
A(B), with $f$ the free energy density and $n_{A(B)0}$ the average density variation
of A(B) component, and $\omega = f - \mu_A n_{A0} - \mu_B n_{B0}$ is the grand
potential density. Given that $\omega = \Omega/V = -P$ with the grand potential
$\Omega$, system volume $V$, and pressure $P$, Eq. (\ref{eq:coexist}) gives the
phase coexisting conditions of equal chemical potentials and equal pressure, i.e.,
\begin{eqnarray}
&& \left. \frac{\partial f}{\partial n_{A0}} \right |_1 (n_{A0_1},n_{B0_1})
  = \left. \frac{\partial f}{\partial n_{A0}} \right |_2 (n_{A0_2},n_{B0_2}) = \mu_0^A,
  \nonumber\\
&& \left. \frac{\partial f}{\partial n_{B0}} \right |_1 (n_{A0_1},n_{B0_1})
  = \left. \frac{\partial f}{\partial n_{B0}} \right |_2 (n_{A0_2},n_{B0_2}) = \mu_0^B, 
  \label{eq:coexist2}\\
&& f_1 - \mu_0^A n_{A0_1} - \mu_0^B n_{B0_1} = f_2 - \mu_0^A n_{A0_2} - \mu_0^B n_{B0_2} = -P, 
  \nonumber
\end{eqnarray}
where $f_1(n_{A0_1},n_{B0_1})$ and $f_2(n_{A0_2},n_{B0_2})$ are the free energy densities
of phase 1 and 2 respectively. 

To obtain the free energy density $f(n_{A0},n_{B0})$ and the corresponding
chemical potentials of A and B components, we use the one-mode approximation for
each ordered phase. The corresponding one-mode expressions assumed in our analytic
calculations are given in the Appendix. For each binary phase the parameters in these
expressions, including the wave number and amplitudes of density field, are determined
from free energy minimization (after substituting the one-mode expressions of $n_A$ and
$n_B$ into the free energy functional Eq. (\ref{eq:F}) and integrating over a unit cell);
from this we then derive the free energy density $f(n_{A0},n_{B0})$ for each phase.
Results for the example of BH phase are presented in the Appendix.

The resulting phase diagram is multi-dimensional, e.g., in the
$\epsilon_A$--$\epsilon_B$--$n_{A0}$--$n_{B0}$ parameter space (with all the other
model parameters fixed). For simplicity, here we consider the A/B symmetric case of
$\epsilon_A=\epsilon_B=\epsilon$, leading to a 3D $\epsilon$--$n_{A0}$--$n_{B0}$ phase
diagram. It would be convenient to calculate the diagrams in two steps: First
identify the stability diagram showing the phase of lowest free energy in each
regime of the parameter space, with phase boundaries determined by the solution
of $f_1(n_{A0},n_{B0}) = f_2(n_{A0},n_{B0})$ for any two phases 1 and 2, and then construct
the corresponding phase diagrams (showing coexistence between two or three phases) based
on Eq. (\ref{eq:coexist2}). Two sample diagrams at $\epsilon=0.1$ and $0.3$ obtained
from our analytic calculations are shown in Figs.~\ref{fig:phase_diagrams}(a) and
\ref{fig:phase_diagrams}(b).

\subsection{Phase diagrams: Direct numerical calculations}

We also calculate the phase diagrams through direct numerical simulations of this
binary PFC model. The dynamic model equation (\ref{eq:nA_nB}) is solved numerically in
the absence of noise terms, starting from random initial conditions across the parameter
space of $(n_{A0}, n_{B0})$ at each specific value of $\epsilon=\epsilon_A=\epsilon_B$.
The corresponding phase at each point of the parameter space is determined by its
steady-state structure. Two of these numerically determined phase diagrams for
$\epsilon=0.1$ and $0.3$ are given in Figs.~\ref{fig:phase_diagrams}(c) and
\ref{fig:phase_diagrams}(d) respectively, showing some quantitatively different
results of phase boundaries as compared to those in Figs.~\ref{fig:phase_diagrams}(a)
and \ref{fig:phase_diagrams}(b) obtained from above analytic calculations. One of
the obvious differences is the much larger regime of BH phase identified from numerical
solutions. In addition, in Fig.~\ref{fig:phase_diagrams}(a) with $\epsilon=0.1$ only
two-phase coexistence is obtained from the analytic results under one-mode
approximation, whereas both two- and three-phase coexistence regions are seen in
Fig.~\ref{fig:phase_diagrams}(b) when $\epsilon=0.3$. This differs from the simulation
results in Figs.~\ref{fig:phase_diagrams}(c) and \ref{fig:phase_diagrams}(d) which show
two- and three-phase coexistence regions for both values of $\epsilon$.

These differences can be attributed to the oversimplified assumptions in the one-mode
expressions of density fields used in the analytic calculations. Generally the A and B
density variation fields are expanded as $n_A = n_{A0} + \sum_j A_j \exp[i ({\bm q}_{A_j}
\cdot {\bm r} + \varphi_{A_j})] + {\rm c.c.}$ and $n_B = n_{B0} + \sum_j B_j
\exp[i ({\bm q}_{B_j} \cdot {\bm r} + \varphi_{B_j})] + {\rm c.c.}$. In the standard
procedure of phase diagram calculation that is followed above in Sec. \ref{sec:analytic},
the amplitudes $A_j$ and $B_j$ are assumed to be real once the wave vectors ${\bm q}_{A_j}$,
${\bm q}_{B_j}$ and phase shifts $\varphi_{A_j}$, $\varphi_{B_j}$ are identified from the
structural symmetry (as given in the Appendix). This procedure has worked well for the
previous single-component and alloy PFC models \cite{Elder02,*Elder04,elder07,
greenwood2011modeling,Ofori-opoku13}. However, the discrepancies shown in
Fig.~\ref{fig:phase_diagrams} between analytic results and numerical solutions
indicate a more complicated scenario for the case of binary sublattice ordering
examined here, particularly regarding the selection of complex phases of A and B
amplitudes. More details for the example of BH structure, including the corresponding
amplitude equations and phase selection, will be presented elsewhere.

\section{Phase transformation, grain nucleation and growth}

Below the melting point crystallites can nucleate homogeneously or heterogeneously
from the supersaturated homogeneous state. In either case, those nuclei will grow
individually until they merge, which usually leads to the formation of topological
defects such as dislocations and grain boundaries in the system. Many factors
(e.g., temperature and average densities) determine the ordered structures
and dynamics arising from those nucleation processes. The emergence of multiple
coexisting phases and the structural transformation between them can also occur
during the system evolution due to the phase coexistence determined in the
phase diagram. We have conducted a series of simulations to examine this nucleation
or phase transformation process, with some sample results given below. Here
the nonconserved dynamics is used, for a better control of the grain growth rate
and the condition of constant flux, i.e.,
\begin{eqnarray}
  && \frac{\partial n_A}{\partial t} = - \frac{\delta \mathcal{F}}{\delta n_A} + \mu_A,
  \nonumber\\
  && \frac{\partial n_B}{\partial t} = - m_B \left ( \frac{\delta \mathcal{F}}{\delta n_B}
  - \mu_B \right ),
\label{eq:nonConsDynam}
\end{eqnarray}
where $\mu_{A(B)}$ is the chemical potential of A(B) component. The process of grain
growth is controlled through tuning the values of $\mu_A$ and $\mu_B$ which emulate
the constant flux condition. No noise terms are added in the above PFC dynamic
equations used in our simulations, given the nature of heterogeneous nucleation
studied here for which noise does not play a crucial role.

\subsection{Nucleation and growth of BH grains}

\begin{figure*}
  \centerline{\includegraphics[width=0.8\textwidth]{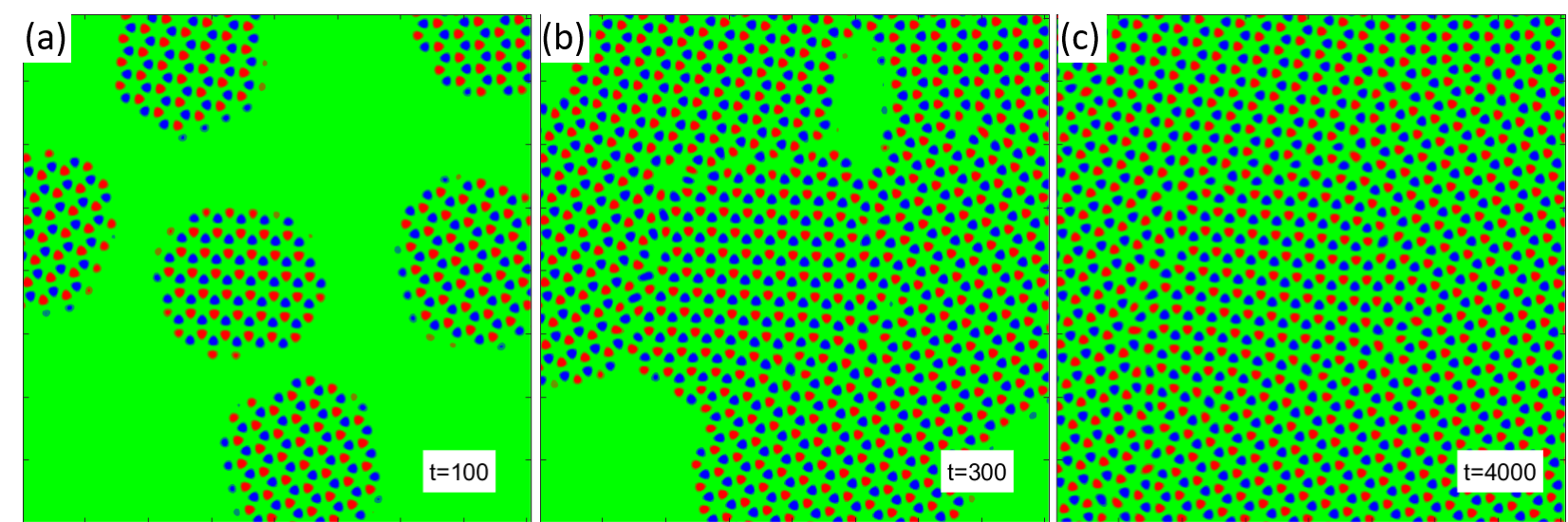}}
  \caption{Grain growth and coalescence process obtained from PFC simulation.
    The nuclei of BH structure grow and impinge to form grain boundaries and a
    polycrystalline system, with a portion of the simulation results shown at
    (a) $t=100$, (b) $t=300$, and (c) $t=4000$.}
  \label{fig:soldification}
\end{figure*}

We first study an example of emergence of binary honeycomb (BH) phase from a homogeneous
(BHom) state, simulating the dynamic process of grain nucleation, individual grain
growth, grain coalescence, and eventually the formation of a polycrystalline state.
Initially twenty circular nuclei of BH structure are placed at random locations in
a simulation box, with randomly assigned different orientations. The nuclei evolve and
grow individually until the grains merge and form a binary honeycomb film. The average
densities for BH and BHom states are set as $n_{A0}=n_{B0}=-0.27$ and $-0.47$,
respectively. The chemical potential $\mu_A = \mu_B$ is set to be $-0.58$, slightly
larger than the corresponding two-phase coexistence value. A portion of the simulation
box is shown in Fig.~\ref{fig:soldification}, giving three snapshots during the
system evolution. Fig.~\ref{fig:soldification}(a) shows the early growth stage of
individual grains before they impinge on each other, where the process of faceting
occurs on the surface of each BH grain which evolves to a hexagon shape. At a later
time stage [Fig.~\ref{fig:soldification}(b)] the impingement of grains has occurred
and coalescence is taking place, which leads to the formation of dislocations and
grain boundaries. For both stages the solidification process is not yet complete
and there is still part of the system that is in the homogeneous state. 
At large enough time the whole system evolves to the ordered state of BH symmetry,
as shown in Fig.~\ref{fig:soldification}(c). The system is polycrystalline, with
grain boundaries separating grains of different orientations.

\subsection{BH-to-BS phase transformation}

\begin{figure*}
\centerline{\includegraphics[width=0.8\textwidth]{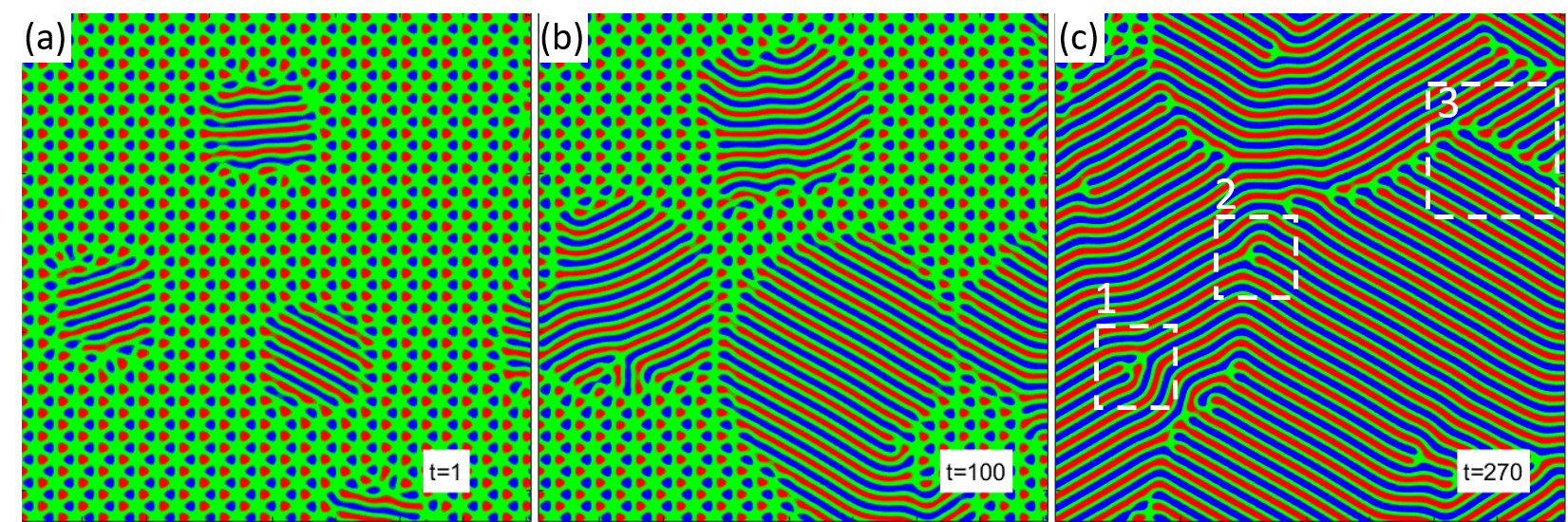}}
\caption{Binary honeycomb (BH) to binary stripe (BS) phase transformation obtained from PFC
  simulation. The system transforms from BH to BS phase as a result of growth and merging 
  of individual BS grains, with a portion of the simulation results shown at (a) $t=1$,
  (b) $t=100$, and (c) $t=270$. Sample defects of disclination, dislocation, and grain
  boundary are indicated in the white dashed boxes with labels 1, 2, and 3 respectively.}
  \label{fig:phasetransform}
\end{figure*}

An example of phase transformation is presented in Fig.~\ref{fig:phasetransform}, showing
the dynamic process of transformation from a binary honeycomb (BH) structure to the binary
stripe (BS) phase. We use a setup similar to the previous section, by initializing twenty BS
nuclei at random locations and orientations in coexistence with the BH matrix. The average
densities for BH and BS states are set as $n_{A0}=n_{B0}=-0.21$ and $-0.0739$, respectively,
and the system chemical potential is chosen as $\mu_A = \mu_B=-0.35$ (above the
corresponding coexistence value). Three snapshots representing different stages of
system evolution are given in Fig.~\ref{fig:phasetransform}, exhibiting BH-BS structural
transformation as a result of the grain growth of BS nuclei and the subsequent grain
coalescence. The individual grain growth rate appears to depend on the initial orientation.
Each single stripe of type A or B grows and connects with the neighboring particles of the
same type. The growth direction of the binary stripes is not restricted to that of the
initial grain, and the fronts could change direction (i.e., curve) during growth. When
the differently oriented grains coalesce some topological defects are formed, including
disclinations, dislocations, and grain boundaries, as indicated in the white dashed boxes
of Fig. \ref{fig:phasetransform}(c).

\subsection{BH-to-ETASB phase transformation}

\begin{figure*}
\centerline{\includegraphics[width=0.8\textwidth]{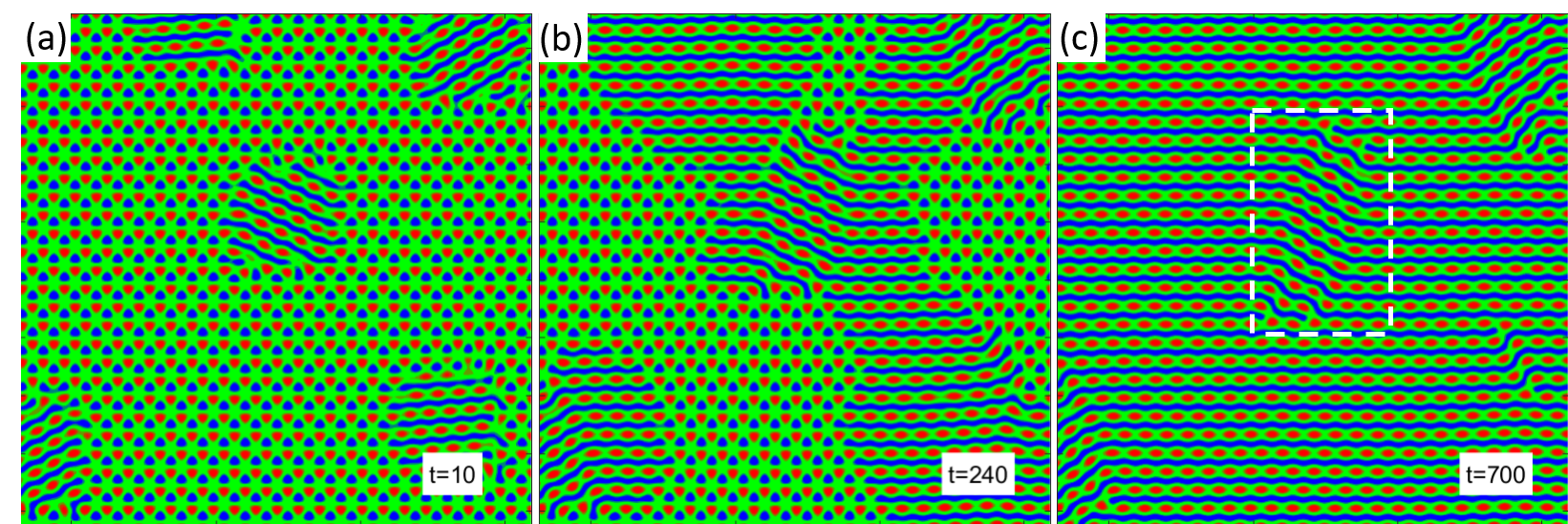}}
\caption{Binary honeycomb (BH) to elongated triangular A \& stripe B (ETASB) phase
  transformation obtained from PFC simulation. Snapshots of the time evolution of
  a portion of the simulated system are shown at (a) $t=10$, (b) $t=240$, and (c) $t=700$.
  A region of kink defects is indicated in (c).}
  \label{fig:phasetransform2}
\end{figure*}

Figure \ref{fig:phasetransform2} presents another example of phase transformation, from
the BH to elongated triangular A \& stripe B (ETASB) phase. The initial setup here is
the same as before, other than the nuclei being of ETASB type as seen in
Fig.~\ref{fig:phasetransform2}(a). The parameters used in the simulation are
$n_{A0}=-0.33$ and $n_{B0}=-0.1427$ for the BH matrix, $n_{A0}=-0.3417$ and $n_{B0}=0.002$
for the ETASB nuclei, $\mu _A=-0.6302$, and $\mu_B=-0.3225$. During the system evolution
the BH structure transforms into ETASB starting at the edges of the growing grain. As seen
in Figs.~\ref{fig:phasetransform2}(b) and \ref{fig:phasetransform2}(c), in this case type
B particles transform from a spatial arrangement of triangular symmetry to stripe. To
accommodate this transformation, type A particle densities transform from a structure of
triangular symmetry to elongated triangular symmetry. If the initial ETASB grain orientation
is around or less than $5^\circ$ with respect to the direction of the surrounding matrix,
then the grain rotates to match the orientation of the matrix. For larger grain orientations,
step defects or kinks are formed. This kink defect acts as a transition between two
different orientations of the merging grains. An example is enclosed by the dashed box
in Fig. \ref{fig:phasetransform2}(c) where the whole system has been transformed to a
defected ETASB state.

\section{Ordered binary structures with competing length scales}

\begin{figure*}
  \centerline{\includegraphics[width=0.9\textwidth]{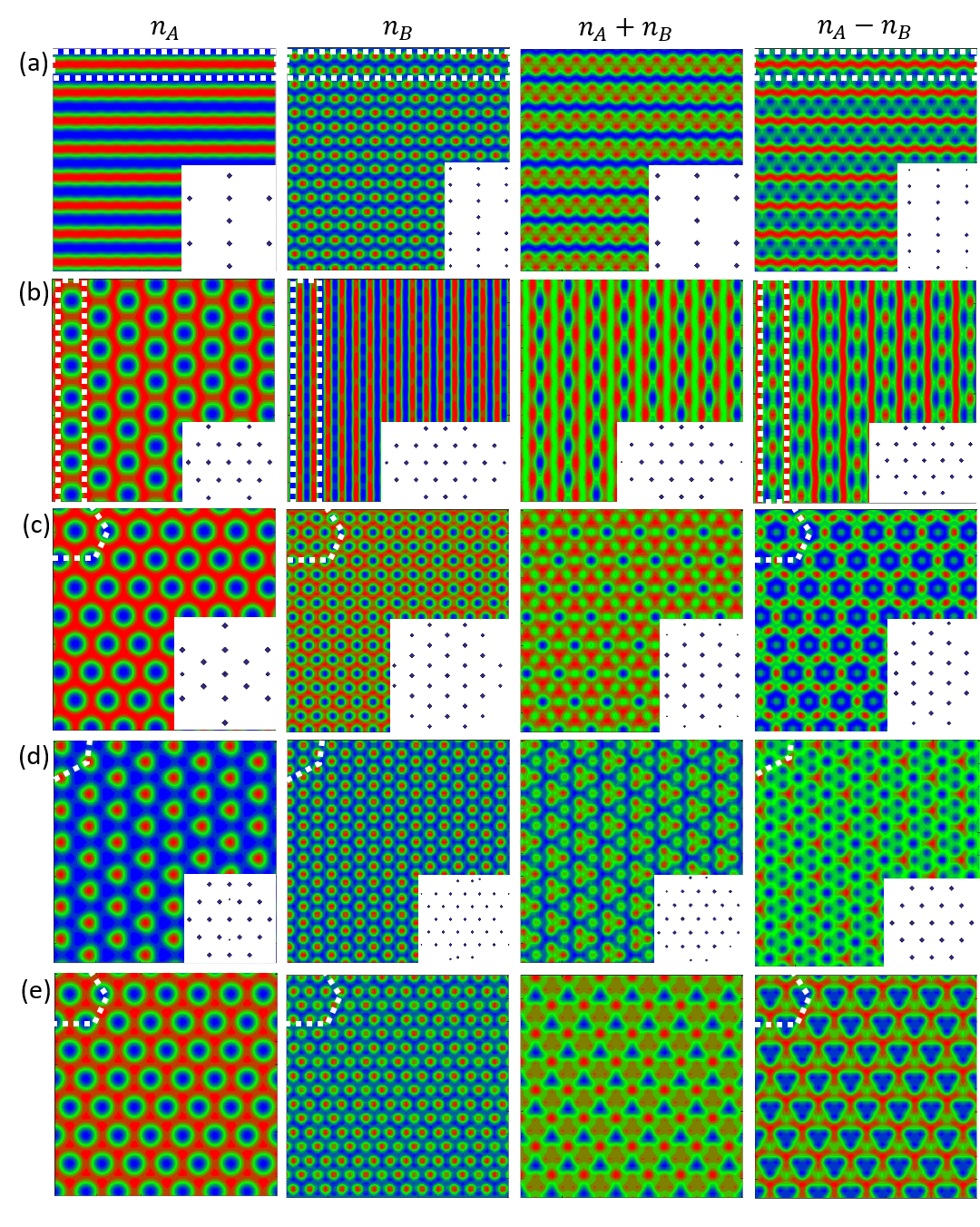}}
  \caption{Some binary ordered structures predicted by PFC simulation, for $q_A=1$ and $q_B=2$,
    and (a) $n_{A0}=0.05$ and $n_{B0}=-0.1$, (b) $n_{A0}=0.35$ and $n_{B0}=0.25$, (c) $n_{A0}=0.25$
    and $n_{B0}=0.4$, (d) $n_{A0}=-0.15$ and $n_{B0}=-0.1$, (e) $n_{A0}=0.3$ and $n_{B0}=-0.1$.
    The first three columns show the spatial distributions of densities $n_A$, $n_B$, and
    $n_A+n_B$ respectively, where red color represents density maximum and blue color
    represents density minimum. The fourth column is for density difference $n_A-n_B$,
    with red color representing density maximum of A component and blue color representing
    density maximum of B component. The diffraction patterns of the corresponding density
    field are shown in the insets.}
  \label{fig:LengthScale1}
\end{figure*}

\begin{figure*}
  \centerline{\includegraphics[width=0.9\textwidth]{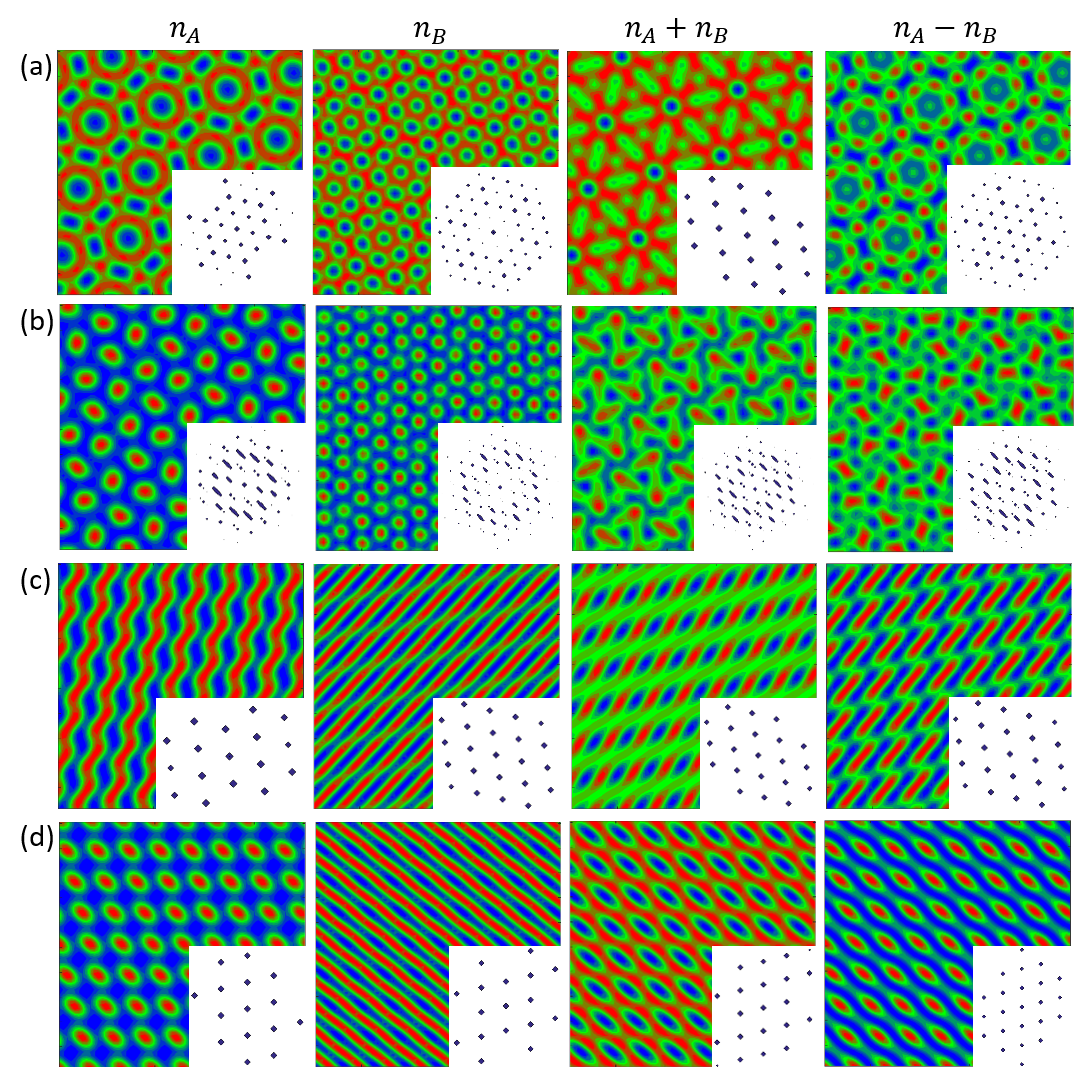}}
  \caption{Some binary ordered structures predicted by PFC simulation, for $q_A=1$ and $q_B=1.62$,
    and (a) $n_{A0}=0.40$ and $n_{B0}=0.40$, (b) $n_{A0}=-0.05$ and $n_{B0}=-0.15$, (c) $n_{A0}=0.25$
    and $n_{B0}=0.25$, and (d) $n_{A0}=0.05$ and $n_{B0}=0.2$. The color scheme and the arrangement
    of columns are the same as that of Fig.~\ref{fig:LengthScale1}.}
  \label{fig:LengthScale2}
\end{figure*}

Two of the key factors controlling the ordering of BiCCs are (i) the coupling and
competition among different length scales, and (ii) the average density variations
of A and B components. The effect of different length scales can be modeled via
changing the ratio between $q_A$ and $q_B$ in the PFC free energy functional
[Eq.~(\ref{eq:F})], i.e., the characteristic wave numbers of the two sublattices.
We then simulate the emergence of the corresponding BiCC structures from the initial
supersaturated homogeneous state, for various values of average density variations
$n_{A0}$ and $n_{B0}$. The system dynamics is governed by Eq.~(\ref{eq:nA_nB}).
Some of the predicted binary ordered structures for two different $q_B/q_A$ ratios
are shown in Figs.~\ref{fig:LengthScale1} and \ref{fig:LengthScale2}, as identified
from our numerical simulations. In all cases the model parameters are chosen as:
$\epsilon_A=\epsilon_B=0.1$, $q_A=1$, $m_B=1$, $\alpha_{AB}=0.5$, $\beta_{AB}=0$,
$g_A=g_B=0.5$, $w=u=0.3$, and $\beta_B=v=1$. 

Results for the length scale ratio $q_B/q_A=2$ are given in Fig.~\ref{fig:LengthScale1},
where the first two columns show the spatial distributions of density fields for each
type of particle, i.e., $n_A$ in the first column and $n_B$ in the second column,
and the third column shows the total density field $n_A+n_B$. In these three columns
the red-colored regions correspond to the maxima of the corresponding density field
and the blue-colored regions represent the minima. Column four presents the density
difference $n_A-n_B$, giving the locations of density maximum for both A component
(red color) and B component (blue color). It is a better representation of the overall
pattern and the A/B coupling as compared to $n_A+n_B$. Five examples given in this
figure [panels (a)--(e)] are obtained from different combinations of average densities
$n_{A0}$ and $n_{B0}$. The individual sublattice structures for $n_A$ and $n_B$ are
of triangular, honeycomb, or stripe type, although additional peaks appear in their
diffraction patterns as compared to the corresponding standard lattice structures,
which can be attributed to the coupling between the two density fields and the two
sublattices.

In Fig. \ref{fig:LengthScale1}(a), $n_A$ exhibits as a modified stripe phase and
$n_B$ as a modified triangular phase with a smaller length scale (given that $q_B/q_A=2$).
The positions of $n_A$ maxima overlay with those of $n_B$ minimum, and for one structural
unit every two rows of $n_B$ maxima correspond to one row of $n_A$ maxima without any
overlaps, as indicated by the white dashed boxes in the first two columns of
Fig.~\ref{fig:LengthScale1}(a). The corresponding superimposed structure is highlighted
by a dashed box in the fourth column representing the density difference $n_A-n_B$.
Similar correspondence between the locations of A and B components can be found in
other types of ordered structures or superlattices. In Fig.~\ref{fig:LengthScale1}(b),
the $n_A$ and $n_B$ distributions show as a modified honeycomb and a modified strip
structure respectively, and every two arrays of $n_B$ maxima correspond to one
array of $n_A$ minima (see the enclosed regions of white dashed boxes inside).
Fig.~\ref{fig:LengthScale1}(c) gives an example where both $n_A$ and $n_B$ are of
honeycomb pattern. In this case, each B honeycomb is enclosed by a larger honeycomb
ring of A component, as can be seen more clearly from the structural unit highlighted
by the white boxes. In Fig.~\ref{fig:LengthScale1}(d), both $n_A$ and $n_B$ are of
modified triangular phase, with smaller lattice spacing for B sublattice. In the
overall pattern of $n_A-n_B$ A particles appear as red forming large-spacing triangular
structure, while B particles (blue) occupy in between, showing as a small-spacing
triangular pattern. The case of honeycomb A and triangular B sublattice structures
is given in Fig.~\ref{fig:LengthScale1}(e). Each unit of the overall binary pattern
is featured by a large honeycomb ring of A component enclosing a smaller triangle
composed of three B particles.

More complicated binary ordered (or quasi-ordered) structures can be obtained when the
length scale ratio $q_B/q_A$ is not an integer. Some sample results are shown in
Fig.~\ref{fig:LengthScale2}, for $q_B/q_A=1.62$. We use similar ways of representing
individual and total density fields and density difference in four columns of the
figure, with the same color scheme for spatial density distribution as above. Since the
B sublattice is of smaller length scale, each of its clusterlike structural unit can be
enclosed inside a larger-scale A unit, as seen in Fig.~\ref{fig:LengthScale2}(a). On the
other hand, the B particles can also orderly distribute within the large spacing of
elongated A particles, as for the example of Fig.~\ref{fig:LengthScale2}(b). Another
possibility is the alternating ordered arrangement of A and B particle clusters,
which forms a superlattice as shown in Fig.~\ref{fig:LengthScale2}(c) and
Fig.~\ref{fig:LengthScale2}(d).

\begin{figure*}
  \centerline{\includegraphics[width=0.9\textwidth]{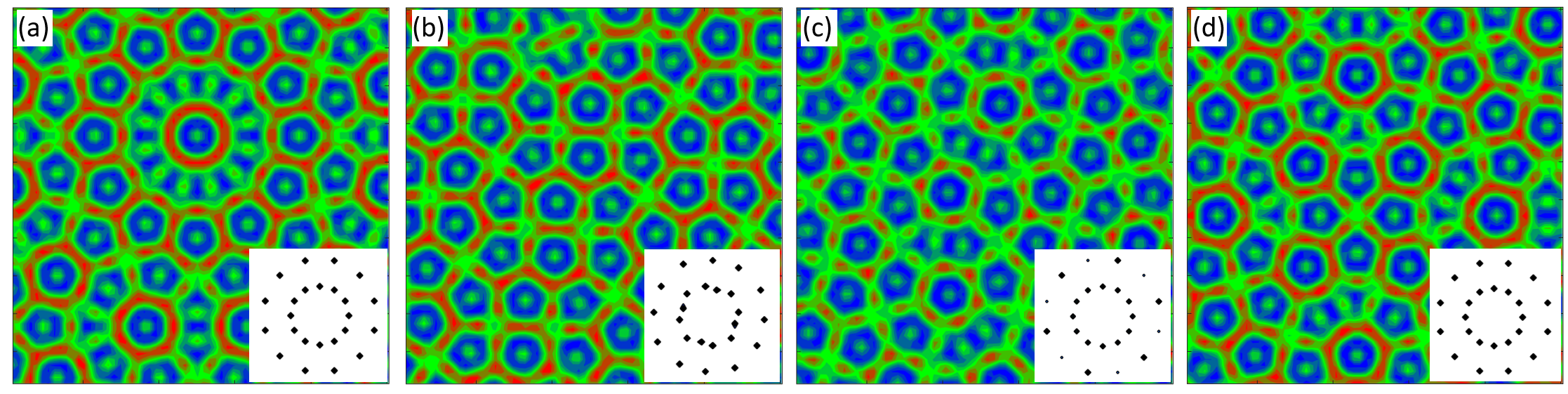}}
  \caption{Sample quasicrystalline patterns obtained from PFC simulation, for
    $q_A=1$ and $q_B=(\sqrt{2}+\sqrt{6})/2$, and (a) $n_{A0}=0.25$ and $n_{B0}=0.33$,
    (b) $n_{A0}=0.28$ and $n_{B0}=0.3$, (c) $n_{A0}=0.27$ and $n_{B0}=0.37$, and
    (d) $n_{A0}=0.3$ and $n_{B0}=0.3$. The structure for density difference $n_A-n_B$
    and the corresponding diffraction pattern are shown in each panel.}
  \label{fig:quasicrystals}
\end{figure*}

Interestingly, when the sublattice length ratio $q_B/q_A$ is irrational and equal to
the characteristic length scale ratio for quasicrystals, e.g., $q_B/q_A=2\cos(\pi/12)
=(\sqrt{2}+\sqrt{6})/2$ for 12-fold symmetry, the corresponding quasicrystalline
structures are expected to emerge, similar to the case of single-component quasicrystals
found in the two-mode PFC modeling \cite{Achim14}. We have obtained some stable
quasicrystalline patterns with binary sublattices through spot checks of simulation
outcomes (both structures and diffraction patterns), with some sample results given
in Fig.~\ref{fig:quasicrystals}. It should be cautioned that the quasicrystalline
structures presented here are actually strained due to the periodic boundary conditions
applied, and more systematic study is needed to further investigate them.

\section{Conclusions}

We have derived a binary PFC model with sublattice ordering based on the classical
dynamic DFT for two-component systems. The model is applied to the study of binary
colloidal crystals, including phase ordering and structural transformations.
Through the control of length scale contrast and coupling between two sublattices
and the tuning of average densities of A and B components, a wide variety of ordered
(or quasicrystalline) structures and superlattices have been generated from the
model. For the simplest case of equal sublattice length scales we identify seven
binary phases, and calculate the corresponding phase diagrams (in the range of
negative average density variations) both analytically and numerically. Much richer
phenomena of binary phase ordering are obtained and predicted for different length
scales of A and B sublattices, which could be the combination of two regular
sublattice ordered structures when the length scale ratio is an integer, or exhibit
as more complex patterns or motifs when the ratio is a noninteger.

The dynamic processes of system evolution and transformation have also been produced
in our simulations. These include grain nucleation, growth, coalescence, and the
formation of topological defects such as grain boundaries, dislocations, disclinations,
and kinks or steps, as demonstrated in the examples of BH grain growth from a homogeneous
state and structural transformation between two ordered phases (e.g., from BH to BS phase
and from BH to ETASB structure) examined here. Our PFC modeling approach and the
underlying mechanisms of scale coupling and competition are of generic nature, and thus
can be straightforwardly extended to the systematic study of different kinds of binary
colloidal crystals including more varieties of binary phases in both two and three
dimensions as well as their ordering and transformation dynamics.

\begin{acknowledgments}
  This work was supported by the National Science Foundation under Grant No. DMR-1609625
  (Z.-F.H.) and DMR-1506634 (K.R.E.).
\end{acknowledgments}

\appendix

\section{One-mode approximation of binary ordered phases}
\label{sec:one-mode}

In the one-mode approximation, the density variation fields $n_A$ and $n_B$ for
various binary 2D ordered phases can be represented by the following:
(i) For the binary honeycomb (BH) phase, due to the triangular ordering of A and B
sublattices that are shifted by ${\bm \delta} = a \hat{y} = (4\pi / 3q) \hat{y}$
with respect to each other, we have
\begin{eqnarray}
n_A &=& n_{A0} + \sum_{j=1}^{3} A_j e^{i {\bm q}_j \cdot {\bm r}} + {\rm c.c.} \nonumber\\
  &=& n_{A0} + 2A_0 \left [ 2\cos \left ( \sqrt{3} q x/2 \right ) 
  \cos \left ( q y/2 \right ) + \cos \left ( q y \right ) \right ], \nonumber\\
n_B &=& n_{B0} + \sum_{j=1}^{3} B_j e^{i {\bm q}_j \cdot ({\bm r} + {\bm \delta})} + {\rm c.c.}
  \nonumber\\
  &=& n_{B0} + 2B_0 \left [ 2\cos \left ( \sqrt{3} q x/2 \right ) 
    \cos \left ( q y/2 + 2\pi/3 \right ) \right. \nonumber\\
    && \left. \qquad\qquad\quad + \cos \left ( q y + 4\pi/3 \right ) \right ].
\label{eq:honeycomb}
\end{eqnarray}
(ii) For the phase of binary stripe (BS), the antiphase of A vs. B field leads to
\begin{eqnarray}
  && n_A = n_{A0} + A_0 \left ( e^{iqy} + {\rm c.c.} \right ) = n_{A0} + 2A_0 \cos(qy),
  \nonumber\\
  && n_B = n_{B0} + B_0 \left [ e^{i(qy+\pi)} + {\rm c.c.} \right ] = n_{B0} - 2B_0 \cos(qy).
  \nonumber\\
\end{eqnarray}
(iii) For the elongated triangular A \& stripe B (ETASB) phase, we assume
\begin{eqnarray}
n_A &=& n_{A0} + 4A_0 \left [ \cos \left ( \sqrt{3} q x/2 \right ) 
\cos \left ( q y/2 \right ) + \cos \left ( q y \right ) \right ], \nonumber\\
n_B &=& n_{B0} + 2B_0 \cos(q y+4\pi/3).
\end{eqnarray}
(iv) The one-mode expression for the phase of triangular A \& honeycomb B (TAHB) is
given by
\begin{eqnarray}
n_A &=& n_{A0} + 2A_0 \left [ 2\cos \left ( \sqrt{3} q x/2 \right ) 
\cos \left ( q y/2 \right ) + \cos \left ( q y \right ) \right ], \nonumber\\
n_B &=& n_{B0} - 2B_0 \left [ 2\cos \left ( \sqrt{3} q x/2 \right ) 
\cos \left ( q y/2 \right ) + \cos \left ( q y \right ) \right ]. \nonumber\\
\end{eqnarray}
The one-mode results for other two variants of ETBSA and TBHA can be expressed
in a similar way.

For the binary honeycomb (BH) phase, substituting the one-mode expression
Eq. (\ref{eq:honeycomb}) into the free energy functional Eq. (\ref{eq:F}) and 
integrating over a cell of $(0 \leq x \leq \sqrt{3} a, 0 \leq y \leq 3a)$ with
$a = 4\pi/3q$, we obtain the free energy density as
\begin{eqnarray}
  &f_{\rm BH} =& f_0 + 3a_1A_0^2 + 3b_1B_0^2 + 4a_2A_0^3 + 4b_2 B_0^3
  + \frac{45}{2} A_0^4 \nonumber\\
  && + \frac{45}{2} v B_0^4 - 3cA_0B_0 - 3wA_0^2B_0 -3uA_0B_0^2,
  \label{eq:f_BH}
\end{eqnarray}
where
\begin{eqnarray}
  &f_0 =& \frac{1}{2} \left ( -\epsilon_A + q_A^4 \right ) n_{A0}^2
  + \frac{1}{2} \left ( -\epsilon_B + \beta_B q_B^4 \right ) n_{B0}^2 \nonumber\\
  && - \frac{1}{3} g_A n_{A0}^3 - \frac{1}{3} g_B n_{B0}^3
  + \frac{1}{4} {n_{A0}}^4 + \frac{1}{4}v {n_{B0}}^4 \nonumber\\
  && + \left ( \alpha_{AB} + \beta_{AB} q_{AB}^4 \right ) n_{A0} n_{B0} \nonumber\\
  && + \frac{1}{2} w n_{A0}^2 n_{B0} + \frac{1}{2} u n_{A0} n_{B0}^2, \label{eq:f0}
\end{eqnarray}
and
\begin{eqnarray}
  && a_1 = -\epsilon_A - 2g_A n_{A0} + 3n_{A0}^2 + wn_{B0} + \left ( q^2 - q_A^2 \right )^2,
  \nonumber\\
  && a_2 = -g_A + 3n_{A0}, \nonumber\\
  && b_1 = -\epsilon_B - 2g_B n_{B0} + 3vn_{B0}^2 + un_{A0}
  + \beta_B \left ( q^2 - q_B^2 \right )^2, \nonumber\\
  && b_2 = -g_B + 3vn_{B0}, \nonumber\\
  && c = \alpha_{AB} + wn_{A0} + un_{B0} + \beta_{AB} \left ( q^2 - q_{AB}^2 \right )^2. 
\label{eq:para}
\end{eqnarray}
The equilibrium state of this binary honeycomb phase is determined by the minimization
of free energy density in terms of wave number $q$ and amplitudes $A_0$ and $B_0$.
Minimizing Eq. (\ref{eq:f_BH}) with respect to $q$ gives
\begin{equation}
q_{\rm eq}^2 = \frac{q_A^2A_0^2 + \beta_B q_B^2 B_0^2 - \beta_{AB} q_{AB}^2 A_0B_0}
{A_0^2+\beta_BB_0^2-\beta_{AB}A_0B_0}.
\end{equation}
In the A/B symmetric case we have $A_0=B_0$ and thus 
$q_{\rm eq}^2 = (q_A^2 + \beta_B q_B^2 - \beta_{AB} q_{AB}^2) / (1+\beta_B-\beta_{AB})$.
Here we consider the simplest scenario of $q_A = q_B = q_{AB} = q_0 = 1$; 
thus $q_{\rm eq} = q_0 = 1$ for any values of $A_0$ and $B_0$,
and Eq. (\ref{eq:para}) becomes
$a_1 = -\epsilon_A - 2g_A n_{A0} + 3n_{A0}^2 + wn_{B0}$, $a_2 = -g_A + 3n_{A0}$,
$b_1 = -\epsilon_B - 2g_B n_{B0} + 3vn_{B0}^2 + un_{A0}$, $b_2 = -g_B + 3vn_{B0}$,
and $c = \alpha_{AB} + wn_{A0} + un_{B0}$.

Minimizing $f_{\rm BH}$ with respect to amplitudes $A_0$ and $B_0$ leads to
\begin{eqnarray}
  &30A_0^3 + 4a_2A_0^2 + 2a_1A_0 - 2wA_0B_0 - uB_0^2 - cB_0 =0,& \nonumber\\
  &30vB_0^3 + 4b_2B_0^2 + 2b_1B_0 - 2uA_0B_0 - wA_0^2 - cA_0 =0.& \nonumber\\
  \label{eq:A0B0}
\end{eqnarray}
The equilibrium amplitudes $A_0^{\rm eq}$ and $B_0^{\rm eq}$ are determined by the solution
of Eq.~(\ref{eq:A0B0}) giving minimum $f_{\rm BH}$, which will then be used in the
calculation of chemical potentials and phase diagrams [see Eq. (\ref{eq:coexist2})].
Similar analysis can be conducted for all other phases based on the assumed expressions
of one-mode approximation given above.

\bibliography{bpfc_2Dsublattice_refs}

\end{document}